\documentclass[12pt]{article}

\usepackage{setspace}
\usepackage{amsmath, amssymb, amsthm,  float,graphicx}
\def\be{\begin{equation}}
\def\ee{\end{equation}}

\def\R{{\mathcal R}}

\def\C{{\mathcal C}}

\def\S{{\mathcal S}}
\def\be{\begin{equation}}
\def\ee{\end{equation}}

\def\a{\alpha}
\def\b{\beta}
\def\g{\gamma}

\def\e{\epsilon}

\def\bg{\bar{g}}
\def\beq{\begin{eqnarray}}\def\eeq{\end{eqnarray}}
\def\ba#1\ea{\begin{align}#1\end{align}}
\def\bg#1\eg{\begin{gather}#1\end{gather}}
\def\bm#1\em{\begin{multline}#1\end{multline}}
\def\bmd#1\emd{\begin{multlined}#1\end{multlined}}
\def\bea{\begin{eqnarray}}
\def\eea{\end{eqnarray}}

\def\a{\alpha}
\def\b{\beta}

\def\e{\epsilon}

\def\g{\gamma}

\def\({\left(}
\def\){\right)}
\def\[{\left[}
\def\]{\right]}

\begin{document}

\title{\bf Entropy functionals and c-theorems\\ from the second law}
\author{Srijit Bhattacharjee$^*$,  Arpan Bhattacharyya$^\dagger$,\\  Sudipta Sarkar$^*$ and  Aninda Sinha$^\dagger$\\
\it $^*$Indian Institute of Technology, Gandhinagar, Gujarat 382424, India.\\
\it $^\dagger$Centre for High Energy Physics,
\it Indian Institute of Science,\\ \it C.V. Raman Avenue, Bangalore 560012, India. \\
}
\date{}
\maketitle
\vskip 2cm
\begin{abstract}{We show that for a general four derivative theory of gravity, only the holographic entanglement entropy functionals obey the second law at linearized order in perturbations. We also derive bounds on the higher curvature couplings in several examples, demanding the validity of the second law for higher order perturbations. For the five dimensional Gauss-Bonnet theory in the context of AdS/CFT, the bound arising from black branes coincides with there being no sound channel instability close to the horizon. Repeating the analysis for topological black holes, the bound coincides with the tensor channel causality constraint (which is responsible for the viscosity bound). Furthermore, we show how to recover the holographic c-theorems in higher curvature theories from similar considerations based on the Raychaudhuri equation. }
\end{abstract}
\newpage
\tableofcontents
\onehalfspacing
%\date{\today}
\vskip 1cm
\section{Introduction}
In the context of AdS/CFT, entanglement entropy of the boundary quantum field theory can be calculated using the Ryu-Takayanagi \cite{Ryu:2006bv, chm} prescription and its generalizations to higher curvature gravity theories. The derivation of this holographic entanglement entropy for two derivative Einstein gravity was proposed by Lewkowycz and Maldacena in \cite{LM}. This derivation allows one to obtain a surface equation (for static situations) for the entangling surface which minimizes an entropy functional. There were attempts to extend this calculation to find the entangling surface equation in higher curvature theories in \cite{bss} and the corresponding entropy functionals were argued to be different from the Wald entropy \cite{Dong:2013qoa,Camps:2013zua,Miao}. In particular, for Lovelock theories the entropy functional is the so-called Jacobson-Myers (JM) \cite{Jacobson:1993xs,Hung} one while for general four derivative theories, it coincides with the Fursaev-Patrushev-Solodukhin (FPS) entropy functional \cite{Fursaev:2013fta}. However, there still exist problems in finding the entangling surface using the Lewkowycz-Maldacena method and there are potential ambiguities related to higher order extrinsic curvature terms in the entropy functionals. Moreover, it leads one to wonder if such entropy functionals could arise independent of AdS/CFT from different and perhaps more fundamental considerations. 

Wald and collaborators \cite{Wald:1993nt, Iyer:1994ys} had established the first law for black hole mechanics for any diffeomorphism invariant theory of gravity and  proposed that the entropy of a black hole is a  Noether charge associated with the Killing isometry generating the horizon. The Wald entropy suffers from various ambiguities \cite{Jacobson:1995uq} and therefore does not provide a unique answer for the horizon entropy. Although, none of these ambiguities contribute to  black hole entropy of a stationary Killing horizon with regular bifurcation surface, the study of the second law of black hole mechanics beyond general relativity (GR) shows that these ambiguities need to be carefully included in the expression of black hole entropy to obtain an increase theorem similar to Hawking area theorem in GR. In fact, for black holes in Lovelock gravity, it is the JM functional which leads to an increase theorem for linearized perturbations \cite{Sarkar:2013swa}. 

Recently, it has been pointed out that the 2nd law for spherically symmetric black holes is satisfied at linear order in perturbations in general four derivative theories of gravity by the holographic entanglement entropy (HEE) functionals \cite{Bhattacharjee:2015yaa}. This has been also generalized beyond spherical symmetry and to generic perturbations up to linear order \cite{Wall:2015raa} and a detailed construction of the entropy functionals has been proposed for any metric theory of gravity such that the linearized second law holds true. For theories with higher curvature terms, the construction produces the HEE functionals. This is a remarkable result since it allows for a way independent of AdS/CFT of deriving entropy functionals--these same entropy functionals therefore find applications in diverse settings. However, what still needs to be addressed is if only the HEE functionals alone do this job. In this paper, we first show that at linear order in perturbations only the HEE functionals can satisfy the second law. We start with an expression for horizon entropy with all possible ambiguous terms relevant for linear order in perturbation. We fix these terms by demanding the validity of the linearized second law and the final entropy coincides with HEE functionals. To be clear, since our analysis is going to be perturbative, we will not be able to rule out higher order extrinsic curvature terms in the functionals (e.g. $O(K^4)$ in the FPS functional)--see section 5. 

We will tackle the question keeping in mind two different motivations. First, we are interested in what happens in the case of asymptotically flat black holes. We will consider two different types of perturbations--the first kind will be due to radially symmetric, slowly falling matter and the second kind being a shear perturbation. We will find that for the Gauss-Bonnet (GB) theory, if we go to second order in perturbations, the second law is automatically satisfied for regular horizons if the GB coupling is positive. If the GB coupling is negative then a certain lower bound on the horizon radius ensures that the second law is satisfied; since this suggests that such black holes cannot be formed from collapsing matter, one can take this as an indication that the negative coupling is disfavoured. If we consider a Ricci-square theory then we find that similar extra conditions on the mass of the black hole may be needed for the second law to hold. Second, we are interested in what happens in the context of AdS/CFT. We will find that for the GB theory, the second order analysis leads to a bound on the Gauss-Bonnet coupling. For black branes, this bound coincides with the absence of sound channel instabilities at the black hole horizon. A similar analysis for the Ricci- square theory bounds the corresponding coupling constant. For GB topological black holes \cite{rgcai}, for the zero mass case we find that the second law bound coincides with the tensor channel causality constraint \cite{Brigante, holoGB}. This bound is what leads to the lower bound on the ratio of shear viscosity to entropy density in the dual plasma and agrees with the lower bound on the $a/c$ ratio in 4d CFTs \cite{hm} with $\mathcal N=1$ supersymmetry. We also investigate quadratic theories with $R_{ab}R^{ab}$ and $R^2$ terms. In these theories, the causality constraints following from \cite{hm} are trivial whereas the second law bound in certain examples we study is nontrivial. 

Using the HEE entropy functionals and the Raychaudhuri equation, we then turn to the question of holographic c-theorems. We will argue that the holographic c-functions found in \cite{ctheorems} arise naturally from such a consideration provided the matter sector satisfies the null energy condition. In \cite{sahakian}, a derivation for the c-functions in Einstein gravity, found in \cite{cthor}, was given using the Raychaudhuri equation. In \cite{cremades} an attempt was made to extend this to higher curvature theories using the Iyer-Wald prescription. Unfortunately the resulting functions do not give the correct central charge (namely the A-type anomaly coefficient in even dimensions). In \cite{ctheorems} it was observed that in all sensible holographic models, which can be constructed by demanding the absence of higher derivative terms in the radial direction (to ameliorate the problem of ghosts) allow for a simple c-function which at the fixed points coincide with the A-type anomaly coefficient in even 
dimensions. This c-function was monotonic under RG flow provided the matter sector satisfied the null energy condition. We will find that using the HEE entropy functionals and considering the Raychaudhuri equation naturally leads to the same c-functions as in \cite{ctheorems}.

The organization of the paper as follows: In the next section we briefly describe the ambiguities in Wald's Noether charge construction. In the section \ref{second-set} set up for proving second law has been described. Next, we determine the coefficients of the ambiguous terms in Wald's construction using linearized second law. In section \ref{beyond}, we go beyond linear order in perturbation to study the GB theory and determine bound on the coupling parameter using AdS black hole solution with different horizon topologies as background. We also consider various other examples like critical gravity theories and put bounds on the couplings in those theories. In section 6, we derive holographic c-functions using the HEE functionals and the Raychaudhuri equation. We end this paper by discussing several follow-up questions. We use $\{-,+,+,,\cdots\}$ signature and set $G=1$ throughout the paper.

\section{Ambiguities in Noether charge method for non-stationary horizons}\label{ambi-wald}
Let us start by describing the geometry of the horizon of a stationary black hole in $D$ dimensions. The event horizon is a null hyper-surface ${\cal H}$ parameterized by a non-affine parameter $t$. The vector field $k^a = (\partial_t)^a$ is tangent to the horizon and obeys non-affine geodesic equation $k^a\nabla_ak^b=\kappa~ k^b$ with $\kappa$ is the surface gravity of the horizon ($a,b,...$ are bulk indices). All $t=$ constant slices are co-dimension $2$ space-like surfaces and foliate the horizon. We construct another auxiliary null normal to the $t=$ constant slices $l^a$, and the inner product between $k^a$ and $l^a$ satisfies $k^a l_a = -1$. The induced metric on any $t =$ constant slice of the horizon is now constructed as $h_{ab} = g_{ab} + 2 k_{(a} l_{b)}$. The horizon binormals are then given by $\epsilon_{ab} = \left( k_a l_b - k_b l_a\right)$.  

The change in the induced metric along these two null directions can be expressed as a sum of trace part and a trace-free symmetric part (assuming null congruences are hypersurface orthogonal therefore have zero twist). The trace part measures the rate of change of the area of the horizon cross section along the null generators which is known to be the expansion and the other part measures the shear of the null geodesic congruences.  We denote $\theta_{k}$ and $\theta_{l}$ to be the two expansion parameters in these two null directions and similarly we have two shears $\sigma_{k}^{ab}$ and $\sigma_{l}^{ab}$. Note that $\theta_{k}$ and $\sigma_{k}^{ab}$ typically vanish on a stationary horizon but $\theta_l$ and $\sigma_l^{ab}$ in general do not vanish. 

As mentioned in the introduction, the entropy of a stationary black hole in any general covariant theory is given by the Noether charge associated with the boost symmetry generating Killing vectors at the horizon \cite{Wald:1993nt,Iyer:1994ys}. The Wald entropy functional for any general covariant Lagrangian $\cal{L}$ in $D$ dimension takes the form: 
\be
S_W =  - 2 \pi \int_\C \frac{ \partial {\cal L}}{\partial R_{abcd} }\epsilon_{ab} \epsilon_{cd} \sqrt{h}\,dA,\label{Wald_Exp}
\ee
where $\C$ denotes any horizon slice and $\sqrt{h}$ is the area element. $R_{abcd}$ is the Riemann tensor of bulk geometry and $dA$ is the area of the $D-2$ dimensional cross-sections of horizon. This formula gives a unique expression for entropy so long one is confined to any stationary slice or to the bifurcation surface of the horizon. However, as pointed out in \cite{Jacobson:1995uq,Iyer:1994ys}, the entropy expression constructed via Noether charge approach has several ambiguities if one tries to apply it for nonstationary slices of the horizon. In fact it turns out, the Wald entropy formula is just one of several possible candidates for the entropy. In \cite{Jacobson:1995uq}, Jacobson, Kang and Myers (JKM) identified three different types of ambiguities which may alter the Wald construction. Among these, the only relevant one for our purpose will be the one which gives us the freedom to add to the Wald entropy a term of the form with arbitrary coefficients:
\be
S_A^{(JKM)}=-2\pi \int_\C X.Y \sqrt{h}\,dA
\label{ambi}
\ee
where the integrand is invariant under a Lorentz boost in a plane orthogonal to an arbitrary horizon slice $\C$  but the terms $X$ and $Y$ are not separately boost invariant \cite{Wall:2015raa,Sarkar:2010xp}.  These ambiguities vanish for Killing horizons but do not necessarily vanish on a nonstationary slice of the horizon. As  a result the entropy functional for a nonstationary horizon may be expressed as\footnote{In the context of entanglement entropy, one typically evaluates this formula on a codimension-2 surface with one timelike $n^{(1)a}$ and one spacelike $n^{(2) a}$ normal such that , $n^{(1)a}=\frac{k^a+l^a}{\sqrt{2}}$ and $n^{(2) a}=\frac{k^{a}-l^{a}}{\sqrt{2}}$ and the extrinsic curvature $\mathcal{K}_{ab}^{(i)}$ for that codimension-2 surface with a induced metric $h_{ab}$ defined as, $\mathcal{K}_{ab}^{(i)}=\frac{1}{2}\mathcal{L}_{n^{(i)}} g_{ab}.$   Trace of this is defined as $\mathcal{K}^{(i)}=\mathcal{K}_{ab}^{(i)}h^{ab}$. Also we have $\mathcal{K}_{i}\mathcal{K}^{i}=-2\theta_{k}\theta_{l}$ and $\mathcal{K}_{(i)ab}\mathcal{K}^{(i)ab}= -2\Big(\frac{\theta_{k}\theta_{l}}{D-2}+\sigma_{k}^{ab}\sigma_{l ab}\Big)$, where $D$ is the dimension of the bulk spacetime. }
\be
\S = - 2\pi \int_\C \sqrt{h} \,dA\, \left[ \frac{\partial L}{\partial R_{abcd}} \epsilon_{ab} \epsilon_{cd}
 -p\, \theta_{k}\theta_{l}- q\, \sigma_k.\sigma_l  \right].
\label{genn1}\ee
\vskip 0.5cm
Here we have used the notation $\sigma_{kab}\sigma^{ab}_l=\sigma_k\sigma_l$. As can be seen from (\ref{genn1}) that the ambiguous terms in the entropy formula involve equal number of $k$ and $l$ subscripts which follows from the fact that these are the only boost invariant combinations that can appear in the entropy functional.   Also on a stationary slice the ambiguity terms vanish and then the expression coincides with the Wald formula (\ref{Wald_Exp}). The First Law of black hole mechanics \cite{Iyer:1994ys, Jacobson:1995uq} doesn't fix the coefficients $p$ and $q$ uniquely. So, to fix the coefficients of these terms one is thus forced to examine whether $\S$ obeys a local increase law. 

 Recently in \cite{Wall:2015raa}, the second law has been shown to hold for any arbitrary higher curvature theories if one allows only linear perturbations to a stationary black hole. This analysis also shows, the entropy functional proposed in the context of Holographic Entanglement Entropy (HEE) \cite{Dong:2013qoa,Camps:2013zua,Fursaev:2013fta} remarkably matches with the $\S$. However, it is not apparent from this analysis\footnote{The criterion in \cite{Wall:2015raa} is sufficient but may not be necessary.}  how just one condition (entropy increases along $t$) fixes two coefficients in the entropy functional! In this paper, we will show explicitly how this fixing happens for curvature squared gravity theories. Note that the unknown coefficients are sitting in front of products of shear and 
expansion terms. Since shear and expansion belong to different irreducible parts of a tensor, they will remain separated even when perturbation is turned on at the linearized level. Consequently we will obtain two independent equations when we demand that entropy increase law holds for every slice of the horizon. This fixes all the coefficients of the entropy functional uniquely for generic quadratic curvature gravity. 

It is important to mention that we have only considered ambiguities which are quadratic order in expansion and shear. We may also have higher powers of such products added to the entropy functional. However, we cannot fix such terms using linearized second law but in curvature squared theories these terms do not enter at linear order \cite{Sarkar:2010xp}. Our analysis matches with the result obtained in  \cite{Bhattacharjee:2015yaa} where the coefficients were fixed for Ricci$^2$ theory using a Vaidya like solution in the linearized increase law. 

\section{Second law set up}\label{second-set}
Let us turn to the apparatus needed to verify the second law of black hole mechanics. The equation of motion for a generic higher curvature metric theory of gravity will be of the form,
\be
G_{ab}\,+\,H_{ab}=8\pi T_{ab}\label{eqm} \,,
\ee
where $G_{ab}$ is the Einstein tensor coming from the Einstein-Hilbert part of the action and $H_{ab}$ is the part coming from higher curvature terms--in theories with a cosmological constant there will also be an additional term proportional to the metric which we can absorb into $G_{ab}$. $T_{ab}$ is the energy momentum tensor which we will assume to obey the Null Energy condition (NEC): $T_{ab} k^a k^b>0$ for some null vector $k^a$. 

We will use the Raychaudhuri equation for null geodesic congruence which describes the evolution of the expansion along the horizon generating parameter $t$. In nonaffine parametrization this looks like\footnote{Interested readers are referred to \cite{Kar, Pois} for discussions on the Raychaudhuri equation.}
\be
\frac{d\theta_k}{dt}=\kappa \theta_k -\frac{\theta_k^2}{D-2}-\sigma_k^2-R_{kk}\,.\label{Ray1}
\ee
Our notation is $A_{ab} k^a k^b = A_{kk}$ and $\sigma_{kab}\sigma^{ab}_k=\sigma^2$. Now, we define an entropy density for any generic higher curvature theory as:
\be
 \S=\frac{1}{4} \int_\C \left(1 + \rho \right)\, \sqrt{h}\,d^{D-2}x, \label{entropyG}
 \ee
where $\rho (t)$ contains contribution from the higher curvature terms including ambiguities. For general relativity, $\rho = 0$. In the stationary limit $\rho$ will coincide with the Wald expression (\ref{Wald_Exp}). Now from this expression the change in entropy per unit area gives us the following expression of generalized expansion $\Theta$,
\be
 \Theta = \frac{d \rho}{dt} + \theta_{k}  \left(1 + \rho \right).
 \label{Theta}\ee
 The evolution of $\Theta$ is governed by the following equation,
\bea
\frac{d\Theta}{dt}- \kappa\Theta=&-&8\pi T_{kk}-\frac{\theta_k^2}{D-2}(1+\rho)-\sigma_{k}^2 (1+\rho)+\theta_k \frac{d\rho}{dt}\nonumber\\&+&H_{kk}+\frac{d^2\rho}{dt^2}-\rho R_{kk}-\kappa\frac{d\rho}{dt}\label{Ray2}\,,
\eea
where we have used the equation of motion (\ref{eqm}) and inserted eq. (\ref{Ray1}). We can now write eq, (\ref{Ray2}) in a convenient form 
\be \label{Ray3}
\frac{d\Theta}{dt}-\kappa \Theta=-8\pi  T_{kk}+E_{kk}\,,
\ee
where 
\be \label{eqn2}
E_{kk}= H_{kk}+\theta_k \frac{d\rho}{dt}-\rho R_{kk}+k^{a}k^{b}\nabla_{a}\nabla_{b} \rho-\left(\frac{\theta_k^2}{D-2} +\sigma_{k}^2\right)(1+\rho)\,.
\ee
We have used $$\frac{d^2\rho}{dt^2}=k^{a}k^{b}\nabla_{a}\nabla_{b} \rho +\kappa \frac{d\rho}{dt}$$ and $$\frac{d}{dt}= k^{a}\nabla_{a}.$$
Next, consider a situation when a stationary black hole is perturbed by some matter flux obeying NEC. The perturbation can be parametrized by some dimensionless parameter $\e$. Note that $T_{kk}$ is linear ($\cal{O}(\e)$) in perturbation and so as $\theta_k$, $\sigma_k$ and $d \rho/dt$. Now, to establish a linearized second law we ignore higher order terms in eq. (\ref{Ray3}). Then, it is easy to see then eq. (\ref{eqn2}) reduces to,
\be
E_{kk}\cong \nabla_k \nabla_k \rho - \rho R_{kk}  +  H_{kk}\label{ekkl}\,.
\ee
We have already mentioned that $T_{kk}$ is of order $\e$, so if the rest of the terms in (\ref{Ray3}) are also collectively of higher order, i.e.,$ E_{kk} \sim \mathcal{O}(\epsilon^2)$, then we obtain

\be
\frac{d \Theta}{d t} - \kappa \Theta= - 8 \pi \,T_{kk}.
\ee
The above equation implies $d \Theta / d t - \kappa \Theta < 0$, on every slice of the horizon. We assume that in the asymptotic future, the horizon again settles down to a stationary state, we must have $\Theta \to 0$ in the future. 
This will imply that $\Theta$ must be positive on every slice prior to the future and as a result the entropy given by (\ref{entropyG}) obeys a local increase law. 

Therefore, to establish the linearized second law, we only need to show that the linear order terms in $E_{kk}$ exactly cancel each other.  Interestingly, this alone will be enough to obtain the values of both the coefficients $p$ and $q$ introduced in the entropy functional (\ref{gen1}). In the next section, we will demonstrate this explicitly for the curvature squared gravity theories. 
 
%%%%%%%%%%%%%%%%%%%%%%%%%%%%%%%%%%%%%

\section{Linearized second law and fixing JKM ambiguities}\label{linear}
 We start with the most general second order higher curvature theory of gravity in five dimensions. The action of such a theory can be expressed as,
\be S = \frac{1}{16 \pi} \int d^5x \sqrt{-g}\left( R- 2\Lambda +\,\alpha\, R^2 +\,\beta \, R_{ab}R^{ab} +\, \gamma\,{\cal L_{GB}}  \right) \label{Flag}
\ee
where ${\cal L_{GB}} = R^2\,-\,4 R_{ab}^2\,+\,R_{abcd}^2$ is the Gauss- Bonnet (GB) combination and $\Lambda$ is the cosmological constant. It is reasonable to assume that such a theory admits a stationary black hole solution as in the case of GR.  We also expect to have a non stationary black hole solution with in-falling matter by perturbing this solution. Such a spacetime will be the counterpart of the Vaidya solution in general relativity for spherically symmetric case and can be expressed as,

\be \label{vd1}
ds^2=-f(r,v) dv^2+2 dv dr + r^2d\Sigma_{3}^2
\ee
$\Sigma_{3}$ can be any three dimensional space with positive, negative or zero curvature. We want to use this solution to investigate the issue of second law of black hole mechanics. Note that, the location of the event horizon $r = r(v)$ for this solution can be obtained by  by solving the following equation,
\be \label{hor}
\dot r= \frac{dr(v)}{dv}=\frac{f(r,v)}{2}.
\ee
with appropriate boundary condition. Note that, $(\,\dot{}\,)$ and $(\,'\,)$  denote respectively derivative with respect to $v$ and $r$. The null generator of the event horizon is given by $k^{a}=\{1,f(r,v)/2,0,0,0\}$ and  the corresponding auxiliary null vector $l_{a}=\{-1,0,0,0,0\}$. The event horizon has nonzero expansion due to the perturbation caused by in falling matter. Next, we will write the entropy associated with the horizon as,
\be
\S = - 2\pi \int_\C \sqrt{h} \,dA\, \left[ \frac{\partial L}{\partial R_{abcd}} \epsilon_{ab} \epsilon_{cd}
 -\tilde p\, \theta_{k}\theta_{l}-\tilde  q\, \sigma_k\sigma_l  \right].
\label{gen1}\ee
We consider several choices of the coefficients $\alpha$, $\beta$ and $\gamma$ etc and study the evolution of this horizon entropy. The aim is to fix the unknown coefficients $\tilde p$ and $\tilde q$ for different gravity theories by demanding the validity of the linearized second law.  

\subsection{Gauss-Bonnet gravity}
We will first study the local increase law for Gauss-Bonnet (GB) case. This corresponds to the choice $\a=\b=0$ in (\ref{Flag}). The action is:
\be
S=\frac{1}{16\pi}\int d^{5}x\sqrt{-g} \Big[R-2\Lambda+\gamma \Big(R_{abcd}R^{abcd}-4R_{ab}R^{ab}+ R^2\Big)+\mathcal{L}_{m}\Big].
\ee
where we have also introduced a matter sector obeying the NEC. $\g$ is a coupling constant of dimension $Mass^{-2}$. The equation of motion for this theory is given by,

\be \label{eom}
G_{ab}+\Lambda g_{ab}+ H_{ab}=8\pi G\, T_{ab}. 
\ee
With,
\begin{eqnarray}
{H}_{ab}\equiv2\g\Bigl[RR_{ab}-2R_{ac}R^c_{~b}
-2R^{cd}R_{acbd} +R_{a}^{~cde}R_{bcde}\Bigr]
-\frac{1}{2}g_{ab} {\cal L}_{GB}.
\end{eqnarray}

Next we start with the expression of the entropy functional (\ref{entropyG}) and evaluate the entropy density for the EGB theory. This is given by sum of Wald entropy density plus the ambiguity terms,

\bea
 \S&=&\frac{1}{4} \int_\C \left(1 + \rho \right)\, \sqrt{h}\,d^3x, \nonumber \\
 &=&\frac{1}{4} \int_\C \left(1 + 2\g(R + 4 R_{kl} - 2 R_{klkl}-p\, \theta_k\theta_l-q\, \sigma_k \sigma_l) \right)\, \sqrt{h}\,d^3x
 \,,\label{eee}\eea
 where $2(R + 4 R_{kl} - 2 R_{klkl})$ is the contribution from the Wald construction. Also $\tilde p= 2 \,\gamma\, p$ and $ \quad \tilde q=2\,\gamma\, q.$  The EGB theory belongs to the general Lovelock class of action functions which give rise to quasi linear equation of motions. It has already been shown in \cite{Sarkar:2013swa} that black holes in all Lovelock theories obey a linearized second law if one uses the JM entropy functional. For EGB gravity, the JM entropy functional is the intrinsic Ricci scalar $(\R)$ of the horizon slice and using the null Gauss-Codazzi equation we can cast $\R$ as,
 
 \be\label{JMGB}
 \R=R + 4R_{kl} - 2 R_{klkl}-{4\over 3}\theta_k\theta_l+2\sigma_k\sigma_l.
 \ee
 
We will now explicitly show below that in EGB gravity the entropy functional which obeys the second law is indeed the JM entropy by fixing the coefficients $p$ and $q$ in (\ref{eee}). 
 
First, we proceed to evaluate the {\it rhs} of eq. (\ref{Ray3}) to study the second law. Note that, if we use the metric (\ref{vd}) with the choice of $d\Sigma_{3}^2$ to be a flat metric then the shear term in (\ref{Ray3}) will vanish identically and $q$ will remain undetermined. This happens because isometries of the metric on a horizon slice essentially coincides with that of a sphere. So we will break the symmetry by adding a cross term and the metric on the horizon slice takes the form,
\be
ds^2=-f(r,v) dv^2+2 dv dr + r^2( dx^2+ dy^2+ dz^2+ \epsilon_{1} h(r,v) dx dy ) .
\ee
We will assume that this shear mode $h(r,v)$ will be balanced by some matter stress tensor still obeying the NEC.  Also we do not  require to find the explicit form of $h(r,v)$ for our analysis. We will calculate $E_{kk}$ order by order in $\epsilon_{1}$ and extract the coefficients of the linear order terms in $\e$ from the evolution equation. Setting those terms to zero will satisfy the linearized second law and in the process $p$ and $q$ will be determined. Below we quote expressions for $\theta's$ and $\sigma's$ for this solution.  
\be\label{exp1}
\theta_{k}=\frac{3 f(r,v) }{2 r(v)}+\mathcal{O}(\epsilon_{1}^2)\,,
\ee
\be\label{shear1}
\theta_{l}=\frac{3}{ r(v)}+\mathcal{O}(\epsilon_{1}^2)\,
\ee
and
\be
\sigma_k\sigma_l=-\frac{\epsilon_1 ^2 \left(r(v)^2 h'(r,v)^2-3 r(v) h(r) h'(r,v)+2 h(r,v)^2\right) f(r,v)}{2 r(v)^6}+\mathcal{O}(\epsilon_{1}^4).
\ee

Now evaluating $E_{kk}$ and extracting the zeroth order terms in $\e_1$   (and linear order in $\e$), we get the following equation,
\be
\frac{\partial^2 f(r,v)}{\partial^2 v}(4-3 p)=0.
\ee
From $\mathcal{O}(\epsilon_{1}^{2})$ terms (which is when $h(r,v)$ makes its first appearance) we get another equation,
\be
2(-6+6p+q) h(r,v)^2+3(2-3p-q)h(r,v)r(v)h'(r,v)+(2+q)r(v)^2h'(r,v)^2=0.
\ee
It is evident that, both of these equations are satisfied if $$p=\frac{4}{3}\,\quad q=-2.$$  This shows that linearized second law fixes all the quadratic ambiguities in entropy functional (\ref{gen1}) uniquely and it is the JM entropy which obeys the linearized second law for GB black holes! Although we have shown this calculation using the 5-dimensional lagrangian for concreteness, the same result holds in any dimension $D>4$.

\subsection{$R_{ab}^2$ theory}

Let us repeat the above analysis for other curvature squared theories. First we take $R_{ab}^2$ theory and we set $\a=\g=0$ and $\beta >0 $ in (\ref{Flag}). We will start with the following entropy functional,

\be\label{e1}
\S=\frac{1}{4}\int d^{3} x \sqrt{h}\Big(1-2\beta(R_{kl}-p\theta_{k}\theta_{l}-q\sigma_k\sigma_l)\Big).
\ee
When $p$ and $q$ are zero then this reduces to the corresponding Wald entropy for the stationary case.  Again proceeding as before we will use the linearized second law to fix these coefficients. Like the Gauss-Bonnet case we get two independent equations and solving them we get, $$p=\frac{1}{2}\,\quad q=0.$$ {As argued in \cite{Bhattacharjee:2015yaa}, the shear part has not contributed to the entropy functional at linear order and the other coefficient of the entropy functional exactly matches with the one which has been shown to obey the linearized second law in \cite{Bhattacharjee:2015yaa}.} \par

\subsection{$R^2$ theory}
We now consider the case $\beta=\g=0$ and $\a >0 $ in (\ref{Flag}) and the entropy functional now reads,
\be\label{e2}
\S=\frac{1}{4}\int d^{3} x \sqrt{h}\Big(1+2\alpha(R-p\theta_{k}\theta_{l}-q\sigma_k.\sigma_l\Big).
\ee
 It has already been argued in \cite{Jacobson:1995uq} that $f(R)$ theory obeys a linearized second law and the Wald entropy functional itself does the job. It is clear, in this case setting $p$ and $q$ equal to zero will correspond to the Wald entropy. If we again repeat the linearized second law analysis we find, $$p=0\,\quad q=0.$$ So this is exactly what is expected from the earlier analysis in \cite{Jacobson:1995uq}.

Therefore, considering all the individual terms in (\ref{Flag}) we have demonstrated that for any general curvature squared theory we can fix the entropy functional completely using only the linearized second law. 
 
\section{Beyond linearized second law}\label{beyond}

In this section, we will consider local entropy increase law beyond linear order which means we are not allowed to neglect the terms in (\ref{Ray3}) which have been thrown away in the previous section. Now we will be considering all the relevant terms in equation (\ref{Ray3}) and will determine the criteria such that second law holds non-perturbatively in the coupling. For this, we will assume a spherically symmetric Vaidya like solution. This means, the metric in (\ref{vd1}) will now have the following form,
\be \label{vd}
ds^2=-f(r,v) dv^2+2 dv dr + r^2d\Omega_{3}^2.
\ee
In fact for EGB gravity one can obtain such a solution just setting the mass in the static spherically symmetric Boulware-Deser \cite{BD} solution to be a function of the advanced time $v$. 

Beyond linearized level, one has to evaluate the full $E_{kk}$ in the evolution equation (\ref{Ray3}). Since the solution is spherically symmetric, shear is identically zero. In fact, for this particular case the evolution equation will be of the form,
\be
\frac{d\Theta}{dt}-\kappa\Theta=-8\pi T_{kk}-\zeta \theta^{2}_k,
\ee
Note that we must have $d\Theta / dt < 0$ to have a local entropy increase law. Also, we have $T_{kk} > 0$ by NEC. Now consider a situation where the stationary black hole is perturbed by some matter flux and we are examining the second law when the matter has already entered into the black hole. In that case, the above evolution equation does not have any contribution from matter stress energy tensor and the evolution will be driven solely by the $\theta_k^2$  term. So we will have a equation of the form,
\be
\frac{d\Theta}{dt}-\kappa\Theta=-\zeta \theta^{2}_k\,.
\ee
In such a situation, if we demand the entropy is increasing, we have to fix the sign of the coefficient of $\theta_k^2$ term. We evaluate the coefficient in the stationary background and impose the condition that overall sign in front of $\theta^{2}_k$ is negative. This will give us a bound on the parameters of the theory under consideration. 
\subsection{Gauss-Bonnet Gravity}
We start with the EGB gravity. The event horizon is now a null surface whose equation is $r = r(v)$. Calculating the r.h.s. of (\ref{Ray3}) for the metric (\ref{vd}) we obtain,
\be
\frac{d\Theta}{dt}-\kappa\Theta=-\zeta \frac{9f(r,v)^2}{4 r(v)^2},
\ee
where we have identified $$\theta_k^2=\frac{9f(r,v)^2}{4 r(v)^2}$$ and introduced $\zeta$ as the coefficient of the $\theta_k^2$ terms. The expression of $\zeta$ is given by,
 \be \label{zeta}
\zeta=\frac{1}{3}\left[1+2\g\left(\, \mathcal{R}-\frac{2}{r(v)}\frac{\partial f(r,v)}{\partial r}\right)\right].\ee\par

For $\g=0$, the coefficient reduces to $1/3$ which matches with GR. $\mathcal{R}$ is the ricci scalar evaluated on the horizon. To satisfy the entropy increase law we now set,

\be\label{cond} \zeta>0.\ee
As discussed earlier, we will evaluate $\zeta$ for different stationary backgrounds and determine bounds on the coefficient $\g$ imposing the condition (\ref{cond}).

\subsubsection{Asymptotically flat case} \label{BD-bnd}
Now we consider the Boulware-Deser (BD) \cite{BD} black hole as the background, for which,
\be
f(r)=1+\frac{r^2}{4\g}\Big[1-\sqrt{1+\g \frac{8 M}{r^4}}\Big].
\ee
Also,
\be \label{eqn3}
r_{h}^{2}+2\g=M
\ee
determines the location of the horizon. Existence of an event horizon demands $r_h^2>0$. In this case horizon topology is a sphere. Now, evaluating $\zeta$ for the above background at the horizon $r =r_{h}$, we get the following inequality
\be 
\frac{M^2+4 M \gamma +36 \gamma^2}{4 \g^2-M^2}<0\,, \label{ineq}
\ee
which leads to  
\begin{eqnarray}
&M>2 |\g|\,\quad if~ M>0\\
&M<-2|\g|\,\quad if~ M<0\,.
\end{eqnarray}
To understand this better, note that we require $M>2\g$ to avoid the naked singularity of the black hole solution for $\g>0$. Thus in this case for a spherically symmetric black hole $\zeta$ will be positive and hence second law will be automatically satisfied. The condition of the validity of the second law is same as that for having a regular event horizon. 

Also, for $\g > 0$, it is possible to make $r_h$ as small as possible by tuning the mass $M$. But when $\g$ is negative (a situation that appears to be disfavoured by string theory, see \cite{BD}, \cite{bms} and references therein), $r_h$ cannot be made arbitrarily small and it would suggest that these black holes cannot be formed continuously from a zero temperature set up. Notice that we could have reached the conclusion without the second law if $M$ is considered to be positive--however, our current argument does not need to make this assumption. Due to this pathology, it would appear that the negative GB coupling case would be ruled out in a theory with no cosmological constant.

%The second condition is immediately ruled out as it corresponds to $r_h^2<0$. Note that this conclusion crucially needs the second law since it is possible to have $M<0$ and $r_h^2>0$ otherwise. Since $M$ corresponds to the ADM mass, we thus have the ADM mass to be positive.
%The condition $M>2\g$ is needed to avoid the naked singularity of the black hole solution for $\g>0$. Thus in this case for a  spherically symmetric black hole $\zeta$ will be positive and hence second law will be automatically satisfied. In this case by tuning $M$ it is possible to make $r_h$ small. When $\g$ is negative (a situation that appears to be disfavoured by string theory, see \cite{BD}, \cite{bms} and references therein), $r_h$ cannot be made arbitrarily small. This appears to be strange since it would suggest that these black holes cannot be formed continuously from a zero temperature set up. Notice that we could have reached the conclusion without the second law if $M$ was positive as in \cite{BD}--however, our current argument does not need to make this assumption. Due to this pathology, it would appear that the negative GB coupling case would be ruled out in a theory with no cosmological constant. 

 \subsubsection{AdS case}
Next we consider the 5-dimensional AdS black hole solution for EGB gravity as the background. The function $f(r)$ now becomes \cite{rgcai},
\be
f(r)=k+\frac{r^2}{4\g}\Big(1-\sqrt{1-\frac{8 \g}{l^2}(1-\left(\frac{r_{0}}{r}\right)^{4}}\Big).
\ee
Here $l$ is the length scale with the cosmological constant ($2\Lambda=-12/l^2$) and $k$ can take values $0,1$ and $-1$ corresponding to planar, spherical or hyperbolic horizons respectively. The intrinsic Ricci scalar on the horizon is $6k/r_h^2$ where, the horizon is at 
\be
r_h^2=\frac{l^2}{2}\left[-k+\left(\frac{4 r_0^4}{l^4}+k^2\left(1-\frac{8 \g}{l^2}\right) \right)^{1/2}\right]\,.
\ee
 First we will consider black brane solution for which $k=0$ and the horizon is planar. The intrinsic curvature of the codimension two slices of the event horizon vanishes. In this case the coefficient $\zeta$ reads,
\be \label{betajm}
\zeta= \frac{1}{3}\left(1-16\frac{\gamma}{l^2}\right).
\ee
Introducing a rescaled coupling $\lambda_{GB} l^2=2\g$ \cite{Brigante, Buchel} we get,
\be
\zeta= \frac{1}{3}(1-8\lambda_{GB}).
\ee
Again demanding positivity of $\zeta$ we get, 
\be
\lambda_{GB}< \frac{1}{8}\,,
\ee
so to satisfy the second law using $\S$ at $\mathcal{O}(\epsilon^2)$ order we need to impose a bound on the GB coupling. 
Incidentally the same has to be imposed to avoid instabilities in the sound channel analysis of the quasinormal mode for dual plasma. It was shown in \cite{Buchel} that when $\lambda_{GB} >\frac{1}{8}$ the Schroedinger potential develops a well which can support unstable quasinormal modes in the sound channel.  It is interesting to see that the second law knows about this instability. \par
Next we consider $k=-1$ case. This will give us a black hole with a hyperbolic horizon. We will also set $r_{0}=0$, which corresponds to the zero mass solution. In this limit we find ,
\be
\zeta=5\sqrt{1-4\lambda_{GB}}-4.
\ee
Demanding $\zeta>0$  we get,
\be
\lambda_{GB} < \frac{9}{100}.
\ee

If we take the limit $r_0/l\rightarrow \infty$ in the expression of $\zeta$ with $k=\pm 1$ (i.e. the case when the horizon sections becomes planar), we recover the bound on GB coupling $\lambda_{GB}$ for the $k=0$ case. However, this is a weaker bound on $\lambda_{GB}$ and the strongest bound on $\lambda_{GB}$ comes from the hyperbolic black hole in the massless limit. This is illustrated in figures 1a and 1b. We have checked that for the $k=1$ case other thermodynamic considerations (e.g., positive entropy, black hole being the correct phase) do not lead to stronger bounds.

\begin{figure}[hb]
\begin{tabular}{cc}
\includegraphics[height=1.8in,width=3.5in ]{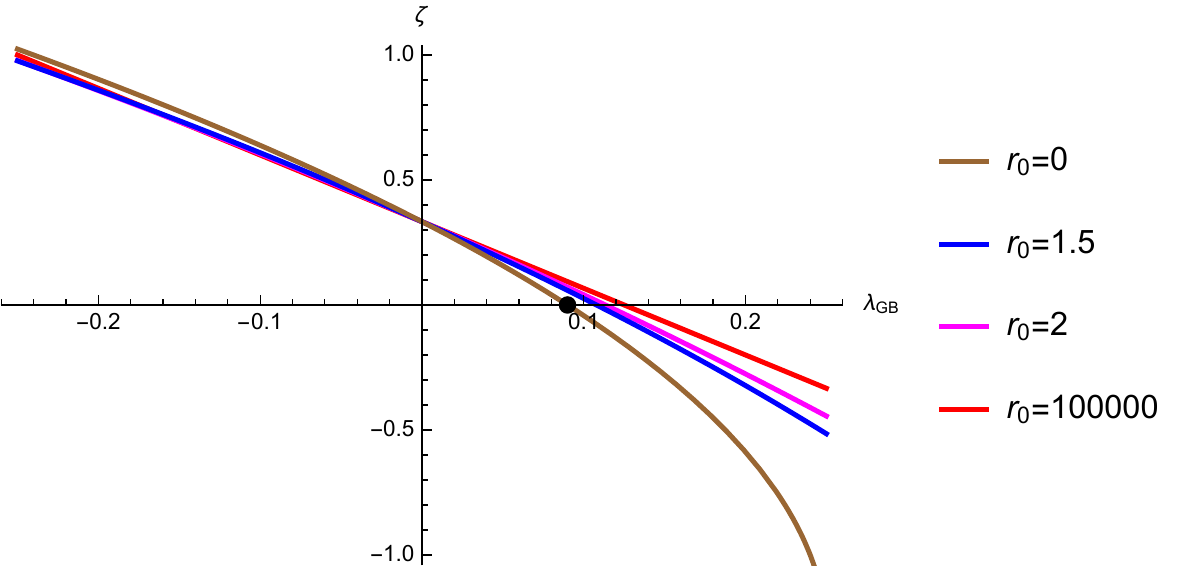} & \includegraphics[height=1.8in,width=3.5in ]{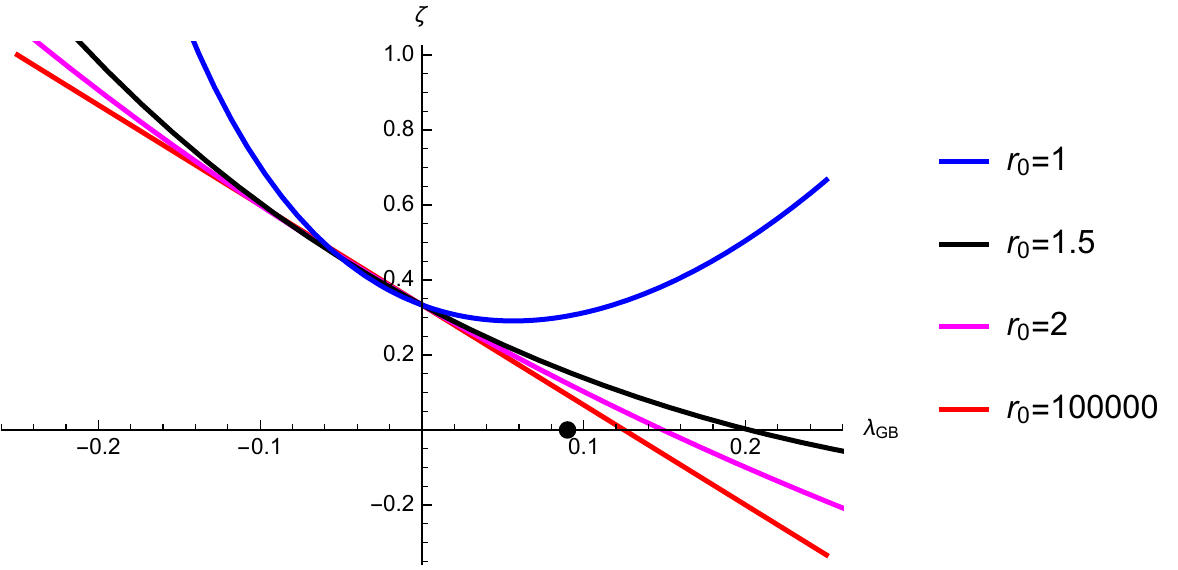} \\
(a) & (b)
\end{tabular}
\caption{(Colour online) (a) Plot of $\zeta$ for different $r_0$ for $k=-1$. (b) Plot of $\zeta$ for different $r_0$ for $k=1$. The black dot denotes $\lambda_{GB}=9/100$. We have set the $l=1$. Note that both the figures shows that $\zeta$ will be positive in all the cases provided $\lambda_	{GB} < 9/100.$}\label{strongbound}
\end{figure}

 \par

Quite curiously we have recovered the tensor channel causality constraint \cite{hm, Brigante, holoGB}\footnote{We can repeat the same analysis for higher dimensions using the normalizations in \cite{holoGB}. Curiously, we find the same bound $\lambda_{GB} < \frac{9}{100}$ which is  stronger than the causality bounds for $d>5$ in \cite{holoGB}.  For flat case ($k=0$), the bound on GB coupling in $d$ dimensions can be expressed as $\lambda_{GB}<1/(2(d-1))$. This means for any dimension the strongest bound arises when the black holes have hyperbolic horizon ($k=-1$) with $r_0=0$. This would also suggest that in the large $d$ limit, $\eta/s\rightarrow 1/4\pi$ contrary to $1/8\pi$ obtained in \cite{holoGB}.}. The bound on $\lambda_{GB}$ in this channel leads to the lower bound on the ratio of shear viscosity ($\eta$) to entropy density ($s$), $4\pi \eta/s\geq 16/25$ in GB holography \cite{Brigante} . The vector and scalar channels also lead to bounds on the coupling from causality constraints. The strongest bound in GB arises from the tensor and scalar channels \cite{holoGB}. Our finding coincides with the tensor channel constraint. It will be interesting to see if the other channels can also be reproduced using the second law analysis. 

 We end this section with one last comment. Instead of putting a bound on the Gauss-Bonnet coupling, one can try to modify the entropy functional  $\S$ itself so that all the $\mathcal{O}(\epsilon^2)$ terms will be canceled. One possible modification is to add a term like $\theta_{k}^2\theta_{l}^2$ which vanishes on the stationary horizon and boost invariant in the null direction. So this type of higher order ambiguous term can occur in the Wald derivation. We modify $\S$ as,
\be
S_{modify}=\frac{1}{4}\int d^{3} x \sqrt{h}(1+2\g\mathcal{R}+ \chi\, \theta_{k}^2\theta_{l}^2).
\ee
$\zeta$ in (\ref{betajm}) will be modified as,
\be \label{modeq}
\zeta=\frac{1}{3}\left[1-\left(\frac{4 \g}{r(v)} \frac{\partial f(r,v)}{\partial r}+ \frac{54 \chi}{r(v)^2}\,  \left(\frac{\partial f(r,v)}{\partial r}\right)^2\right)\right]
\ee
So we can adjust $\chi$ such that the $\zeta$ vanishes and second law automatically holds up to $\mathcal{O}(\epsilon^2)$. One may also can add a term like $\nabla_{i}\theta_{k}\nabla^{i}\theta_{l}$ at this order where $\nabla_{i}$ is the covariant derivative with respect to the surface index. But these additional terms will generate further higher order terms starting from $\mathcal{O}(\epsilon^3).$ So one will need to add more terms to cancel them. It might possible to do this recursively. 
However, the coefficients of these terms will also depend on the background data.  So it will be prudent not to modify the entropy functional. We have put a bound on the coupling instead of modifying the entropy functional at $\mathcal{O}(\epsilon^2)$ order such that second law continues to hold up to this order.  It will be interesting to investigate what happens to this bound on $\lambda_{GB}$ when one goes beyond $\mathcal{O}(\epsilon^2)$ order.

\subsection{$R_{ab}^2$ theory}

\subsubsection{Asymptotically flat space}
First consider the asymptotically flat black hole with spherically symmetric horizon in 5d. In this case,
\be
\zeta={1\over 3}\Big(1+\frac{8 \beta}{ r(v)^2}+ \frac{15}{2} \beta  f''(r,v)+\frac{3   }{2 r(v)}\beta f'(r,v)-\frac{25}{2 r(v)^2} \beta  f(r,v)\Big)
\ee
Now, to extract a bound,  we need an explicit form of a spherically symmetric static black hole solution in $Ricci^2$ theory. We are not aware of any such solution except the 5d Schwarzschild solution of GR which is also a solution of $Ricci^2$ theory. So, as an example we take such a solution as the background and write,
\be
f(r)=1-\frac{r_h^2}{r^2}.
\ee
Then the validity of the second law gives us the condition, 
\be
\frac{r_h^2 }{34}> \beta.
\ee
When $\beta<0$ this is automatically satisfied whenever a horizon exist whereas for $\beta>0$, a lower bound on the horizon size ensures the validity of the second law. \\

Repeating this same analysis for 4d with the metric function $f(r,v)=1- (r_h/ r)$ we get ,
\be
\frac{r_h^2}{18}>\beta,
\ee Therefore, in both cases, there is a minimum horizon radius for the second law to hold when $\beta>0$. Again as discussed in \ref{BD-bnd}, minimum $r_h$ scenario would appear pathological and may be taken as a reason to disfavour the $\beta>0$ sign.

\subsubsection{AdS case}
We repeat the same analysis of the previous section for $R_{ab}^2$ theory but with the background as an asymptotically AdS black brane in 5d. The metric function is then given by
 \be
 f(r)=\frac{r^2}{l^2} f_{\infty}\Big(1-(r_{h}/{r})^4\Big )
 \ee
where $f_{\infty}=\big (\frac{l}{l_{AdS}}\big)^2$ and it satisfies the equation,
\be
%f_{\infty} (2 f_{\infty} \lambda_{2}-3)+3=0.
1-f_\infty+\frac{2}{3}\lambda_2 f_\infty^2=0\,.
\ee
In this case we obtain,
\be
\zeta=\Big(\frac{1}{3}+\frac{5}{2} \beta  f''(r,v)+\frac{\beta  f'(r,v)}{2 r(v)}-\frac{25 \beta  f(r,v)}{6 r(v)^2}\Big)\,,
\ee
 Define $\lambda_{2} l^2=2\beta $   and after evaluating on the horizon with the above metric function as background, we get 
\be 
\zeta=\sqrt{9-24 \lambda_{2}}-\frac{8}{3}.
\ee
From this, the bound becomes,
\be
\lambda_{2} <\frac{17}{216}.
\ee
In this case if we repeat out analysis for a zero mass hyperbolic black hole  we obtain a lower bound on $\lambda_{2}$. Combining them we get, 
\be
-\frac{29}{243} <\lambda_{2} <\frac{17}{216}.
\ee 
If we repeat the same analysis for 4d and obtain the following bound,
$
\lambda_{2} <\frac{2}{15}.
$
No analog of such a bound in the context of AdS/CFT is known in this case. This analysis also suggests that the second law bound arising from hyperbolic horizons is not necessarily connected to the causality constraints. The reason is that for these theories, in the analysis of \cite{hm}, we only get $c_T>0$. 
\subsection{Critical gravity}
We will now consider bounds arising from the second law in quadratic curvature theories with a negative cosmological constant, which only involve terms like $R_{ab}R^{ab}$ and $R^2$. In particular, we will focus on the case of critical gravity  \cite{pope,others}. For any theory with only Ricci$^2$ or $R^2$, the positive energy condition of Hofman-Maldacena \cite{hm} does not yield any constraint (see \cite{anoms} for general expressions) except for the positivity of the two point function of stress tensors which follows from unitarity. In all the examples below, the positivity of black hole entropy leads to the same condition arising from demanding $c_T>0$ where $c_T$ is the coefficient in the two point function of CFT stress tensors. The second law constraint will lead to further conditions as we will see.
\subsubsection{D=4}
The action for the 4d version of critical gravity is given by,
\be S = \frac{1}{16\pi}\int d^4x \sqrt{-g}\left( R+\frac{6}{l^2} +\alpha l^2\left(R_{ab}R^{ab}-\frac{1}{3}R^2\right) \right) \label{clag}
\ee
In general this theory posses a massive spin-2 mode in addition to the usual massless spin-2 degree of freedom. It has been shown in \cite{pope}, at $\alpha=-1/2$, which is called the critical point, the additional massive spin-2 mode becomes massless. Also, at the critical point the entropy of the Schwarzschild-AdS black hole becomes zero. 

We will now use the second law to put a bound on the coupling $\alpha$. Proceeding as before at the second order we get,
\be
\zeta=\Big(\frac{1}{2}-\frac{2\, \alpha \, l^2 \, f'(r,v)}{3\, r(v)}+\frac{2}{3} \alpha\, l^2  f''(r,v)-\frac{2}{3} \,\alpha\, l^2\,  r(v) f'''(r,v)\Big).
\ee
In this case, we take the background as,
\be
 f(r)=\frac{r^2}{l^2} \Big(1-\Big(\frac{r_{h}}{r}\Big)^3\Big ).
 \ee 
 From this we get the bound on the coupling as,
\be \label{b1}
\alpha \leq \frac{1}{12}.
\ee
Also, the black hole entropy has to be positive--if it was zero then it must remain zero at all times. From this we have the following condition,
\be
1+2\,\alpha >0.
\ee
So,
\be \label{b2}
\alpha > -\frac{1}{2}.
\ee
Combining (\ref{b1}) and (\ref{b2}) we get,
\be \label{b3}
-\frac{1}{2} <\alpha \leq \frac{1}{12}.
\ee
 Now from the  AdS/CFT perspective it is natural to assume the positivity of the coefficient $c_{T}$ of two point correlator of boundary stress tensor.  That will give exactly the same condition as in (\ref{b2}) which is weaker than (\ref{b3}).  Repeating the same analysis for a zero mass hyperbolic black hole we did not any further bounds.

\subsubsection{D=5}
Next we consider the critical gravity theory in 5d. The action is given by,

\be S = \frac{1}{16\pi}\int d^5x \sqrt{-g}\left( R+\frac{12}{l^2} +\alpha\, l^2\left(R_{ab}R^{ab}-\frac{5}{16}R^2\right) \right) \label{clag}
\ee
The critical point now corresponds to $\alpha=-5/27$ \cite{pope}. Proceeding as before at the second order we get,
\be
\zeta=\Big(\frac{1}{3}+\frac{\alpha\,l^2  f'(r,v)}{2 r(v)}+\frac{5}{24} \alpha\,l^2  f''(r,v)-\frac{5}{12}\, \alpha \,l^2 r(v)\, f'''(r,v)\Big).
\ee
In this case, the background solution is taken as,
\be
 f(r)=\frac{r^2}{l^2} f_{\infty}\Big(1-\Big(\frac{r_{h}}{r}\Big)^4\Big ),
 \ee 
where the quantity $f_{\infty}$ satisfies the equation, 
$
1-f_{\infty}-\frac{3}{4}  \alpha f_{\infty}^2=0.
$
The second law thus leads to,
\be \label{b4}
-\frac{1}{3}\leq \alpha \leq \frac{109}{2809}.
\ee
Again demanding the entropy to be positive we get,
$
1+\frac{9}{2} f_{\infty}\alpha>0,
$
from which
$
\alpha > -\frac{5 }{27}.
$
Thus we have,
\be \label{b6}
-\frac{5 }{27} < \alpha \leq \frac{109 }{2809}.
\ee
  If we repeat our analysis for a zero mass hyperbolic black hole, the resulting bound we obtain is weaker than this.

\subsubsection{D=3}
We end this section by exploring the case of NMG theory in 3d. 
The action is given by,

\be S = \frac{1}{16\pi}\int d^3x \sqrt{-g}\left( R+\frac{2}{l^2} +\alpha\, l^2\left(R_{ab}R^{ab}-\frac{3}{8}R^2\right) \right) \label{clag}
\ee
The critical point is now at $\alpha=-3.$ For this case,
\be
\zeta=\Big(1-\frac{7 \alpha\, l^2  f'(r,v)}{2 r(v)}+\frac{9}{4} \alpha\, l^2  f''(r,v)-\frac{3}{2} \alpha\, l^2  r(v) f'''(r,v)\Big).
\ee
The background solution is taken as,
\be
 f(r)=\frac{r^2}{l^2} f_{\infty}\Big(1-\Big(\frac{r_{h}}{r}\Big)^2\Big ),
 \ee 
where the quantity $f_{\infty}$ satisfies $
1-f_{\infty}+ \alpha f_{\infty}^2=0.
$
Then we get the following bound,
\be \label{b7}
\alpha \leq \frac{9}{25}.
\ee
Demanding the positivity of the black hole entropy we get,
$
1+\frac{ f_{\infty} \,\alpha}{2}>0.
$
This gives, 
$
-3 < \alpha < 1.
$
Thus we have,
\be
-3 < \alpha \leq \frac{9}{25}.
\ee
 In this case, even if we consider a zero mass hyperbolic black hole we obtain the same bound.
%For both the examples considered above the critical point in holography pertains to setting the coefficient $c_T$ in the two point functions of the boundary stress tensor to zero. In the EGB theory the bound $\lambda<1/8$ prevents $c_T=0$ from happening. However, in the $\alpha R_{ab}R^{ab}+\beta R^2$ example it appears that the second law analysis using the perturbation due to spherically symmetric in-falling matter does not rule out the $c_T=0$ possibility--in fact in both the 4d and 5d examples, $c_T$ could also turn negative. It could be that more generic perturbations will be sensitive to such pathologies.

 \section{Holographic $c$-functions from the entropy evolution equation}
 In this section we will  construct  a holographic $c$-function using the FPS entropy functional for general curvature squared  theories.  Here we will again use the evolution equation for $\Theta$ (\ref{Ray3}) and in the process we will have a geometric derivation for  the $c$-function.  \par
 For unitary Lorentz invariant theory one can define a function in the  space of couplings which decreases monotonically from UV to IR, parametrizing  the RG flow. It coincides with the central charges of the corresponding  CFTs at the UV and IR fixed point. In $1+1$ and $3+1$ dimensions this  has been established rigorously \cite{Zam,Kom}. Now from the holographic point view this $c$ function plays a pivotal role in understanding nature of RG flow.  A  proposal for the $c$ function for arbitrary theories of gravity has been given in \cite{ctheorems,ctheorems2, quasitop} for arbitrary dimensions.  In  this section we will use the Raychaudhuri equation as discussed in the previous sections and derive a $c$ function for a general curvature squared theory. For general higher curvature theories, we do not expect a c-theorem as there are bound to be problems with unitarity in such theories. Thus some condition on the couplings, which presumably only explore a subset of conditions leading to unitarty as in \cite{ctheorems}, is expected. This is what we will find as well. At the UV and IR  fixed points this function will coincide with the A-type anomaly coefficient for the curvature squared theory and  it will also have to be a monotonically decreasing quantity along the RG  flow. To prove the monotonicity we will use Raychaudhuri equation. The basic setup is similar as before.
 \subsection{Setup}
  To derive a $c$ function purely in terms of geometric quantities we will first start with a specific example. We will consider  domain wall type of geometry in $5$ dimensions, the metric for which is given below,
  \be
  ds^2=dr^2+e^{2 A(r)}(-d\tau^2+dx^2+dy^2+ dz^2).
  \ee
 The spacetime can be foliated by surfaces of codimension-2 at every constant time slice.  The induced metric on such surfaces  is given by,
 \be
 ds^{2}_{surface}=e^{2 A(r)}(dx^2+dy^2+dz^2).
 \ee
 Next we will consider a hypersurface orthogonal null congruence.  The tangent vector for this congruences is given by,
 \be
 k^{a}=(\partial_{\lambda})^a=\{-e^{-A(r)},e^{-2 A(r)},0,0,0\}.
 \ee
 It satisfies $
 k^{a}\nabla_{a}k^{b}=0$
  and 
 $k_{a}k^{a}=0.
 $
 We can construct another auxiliary null vector $l^a$ as,
 \be
 l^{a}=\{\frac{e^{A(r)}}{2},\frac{1}{2},0,0,0 \}
\ee
such that, $k_{a}l^{a}=-1.$ Note that in this case $k^a$ and $l^a$ are not any kind of horizon null  generators as compared to the previous sections. They simply correspond to a null congruence that converges along the light sheet projected out of the codimension-2 surface under consideration.   Also in this section we will use affine parametrization. 

\subsection{Holographic c-functions in four derivative gravity}
Next we will start with a guess for the $c$ function for curvature squared theory.  The action for this theory is given by,
\be
S=\frac{1}{16\pi }\int d^{4} x\Big[R+\a R^2+\beta R_{ab}R^{ab}+\gamma (R^2-4 R_{ab}R^{ab}+R_{abcd}R^{abcd})\Big].
\ee
The corresponding entropy functional is the FPS functional.
\be
S_{FPS}=\frac{1}{4}\int d^{3} x \sqrt{h}(1+\rho)= \frac{1}{4 } \int d^{3} x \sqrt{h} \Big(1+2\Big [\alpha R-\beta(R_{kl}-\frac{1}{2}\theta_{k}\theta_{l})+\gamma\mathcal{R}\Big]\Big).
\ee
$\mathcal{R}$ is the 3 dimensional Ricci scalar  intrinsic to the surface. 
 Now we will start with the following candidate,
 \be
 c(\lambda)=\frac{\Theta}{\sqrt{h}\theta_k^4}
 \ee
 where as usual, $\Theta= \theta_k(1+\rho)+\frac{d\rho}{d\lambda}$ and $\lambda$ is the affine parameter. This appears to be a natural guess since for the Einstein theory it goes over to the c-function in \cite{cthor}. Then we have to show two things, 
\begin{enumerate}
 \item $c(\lambda)$ is monotonically decreasing  under the flow along the null geodesic congruences. That is we have to check  the sign of $\frac{dc(\lambda)}{d\lambda}.$ \par
 \item Further it has to coincide with the correct central charge (A-type anomaly in four dimensions) at the fixed points.  \end{enumerate}
 Now,
 \begin{align}
 \begin{split}
 \frac{ d c(\lambda)}{d\lambda} &= \frac{1}{\sqrt{h}\theta_k^4}\frac{d \Theta}{d\lambda}-\frac{ \Theta }{(\sqrt{h}\theta_k^4)^2}\frac{d}{d\lambda}(\sqrt{h}\theta_k^4)\\ &=\frac{1}{\sqrt{h}\theta_k^4}\frac{d \Theta}{d\lambda}-\frac{\Theta }{\sqrt{h}\theta_k^5}(\theta_k^2+4 \frac{d\theta_k}{d\lambda})
 \end{split}
 \end{align}
 using  $\frac{d\sqrt{h}}{d\lambda}=\sqrt{h}\theta_k.$ Next we will replace $\frac{d\Theta}{d\lambda}$ by (\ref{Ray2}) with $\kappa=0$ and $\frac{d\theta_{k}}{d\lambda}$ by Raychaudhuri equation (\ref{Ray1}). 
 We obtain,
  \be
  \frac{dc(\lambda)}{d\lambda}=\frac{24 \pi G T_{kk}}{\sqrt{h}\theta_k^4}+ E_{kk},
  \ee
  where, 
  \begin{align}
  \begin{split}
  E_{kk}=&\frac{e^{-A(r)} (16 \alpha +5 \beta)^2}{243 A'(r)^5}\Big((2 A^{(3)}(r)+15 A'(r)^3+16 A'(r) A''(r)) (15 A'(r)^4+2 (A^{(4)}(r)+\\&8 A''(r)^2)+8 A^{(3)}(r) A'(r)+6 A'(r)^2 A''(r))\Big)+\frac{e^{-A(r)} (16 \alpha +5 \beta )}{486 A'(r)^4}\Big(-24 (A^{(4)}(r)\\&+8 A''(r)^2 )+60 A'(r)^4 (4 (10 \alpha +2 \beta +3 \gamma ) A''(r)-3)+A'(r)^2 A''(r)\\& (256 (10 \alpha +2 \beta +3 \gamma ) A''(r)-207 )+8 A^{(3)}(r) A'(r) (4 (10 \alpha +2 \beta +3 \gamma ) A''(r)-15)\Big)\\&-\frac{4 e^{-A(r)} (10 \alpha +2 \beta +3 \gamma ) A''(r)}{27 A'(r)^2}.
  \end{split}
  \end{align}
   Also,
   \begin{align}
   \begin{split}\label{dcdt}
   \frac{dc(\lambda)}{d\lambda}&=\frac{ e^{-A(r)} (16\alpha+5\beta)}{162 A'(r)^5}\Big(-2 A^{(4)}(r) A'(r)+8 A^{(3)}(r) A''(r)-16 A^{(3)}(r) A'(r)^2\\&+15 A'(r)^3 A''(r)+48 A'(r) A''(r)^2\Big)-\frac{e^{-A(r)} A''(r)}{9 A'(r)^4}
   \end{split}
   \end{align}
   and 
   \be
   c(\lambda)=\frac{2 (16 \alpha +5 \beta ) A^{(3)}(r)+A'(r) \left((16 \alpha +5 \beta ) \left(16 A''(r)+15 A'(r)^2\right)-6\right)}{162 A'(r)^4}.
   \ee
Notice that in (\ref{dcdt}) the term proportional to $16\alpha+5\beta$ has no possibility of having a definite sign and it does not cancel with terms in $E_{kk}$ so it is natural to set the coefficient to zero. 
  Now $\sqrt{h}\theta_{k}^4$ is positive and $T_{kk}$ using the null energy condition is positive.  With $ 16\alpha+5\beta=0$,  we find
  \be
 E_{kk}= \frac{d}{d\lambda}\Big(-\frac{4 (10 \alpha +2 \beta +3 \gamma )}{27 A'(r)}\Big).
  \ee
   where we have used $\frac{d}{d\lambda}= k^{a}\nabla_{a}.$
    So we will have now,
   \be
  \frac{dc(\lambda)}{d\lambda}=\frac{24 \pi G T_{kk}}{\sqrt{h}\theta_k^4}+\frac{d}{d\lambda}\Big(-\frac{4 (10 \alpha +2 \beta +3 \gamma )}{27 A'(r)}\Big),
  \ee
  Then we can define a effective quantity,
  \be
  \bar c(\lambda)= c(\lambda)+\frac{4 (10 \alpha +2 \beta +3 \gamma )}{27 A'(r)}
  \ee
  such that,
 \be
  \frac{d\bar c(\lambda)}{d\lambda}=\frac{24 \pi G T_{kk}}{\sqrt{h}\theta_k^4} \geq 0,
  \ee
Next evaluating $\bar c(\lambda)$ and multiplying both side by a factor of  $-\frac{1}{27}$ we get,
\be
\frac{d}{d\lambda}\Big(\frac{1-4 (10 \alpha +2 \beta +3 \gamma ) A'(r)^2}{A'(r)^3}\Big) \leq 0.
\ee
 Now  in terms of the radial coordinate $r$ this becomes,
 \be
(k^{r}\partial_{r} + k^{\tau}\partial_{\tau} )\Big(\frac{1-4 (10 \alpha +2 \beta +3 \gamma ) A'(r)^2}{A'(r)^3}\Big) \leq 0.
\ee
 From this,
  \be
  e^{-2 A(r)} \frac{d}{dr } \Big(\frac{1-4 (10 \alpha +2 \beta +3 \gamma ) A'(r)^2}{A'(r)^3}\Big) \geq0.
\ee
 So , $$ \bar c(r)=\frac{1-4 (10 \alpha +2 \beta +3 \gamma ) A'(r)^2}{A'(r)^3}$$ is a monotonically decreasing quantity along the RG flow, where the RG scale is determined by the radial coordinate $r$. At the fixed point the geometry is AdS and hence $$A(r)\equiv \frac{r}{l_{ads}}.$$ The $\bar c(r)$ coincides (upto an inconsequential overall factor) with the A-type anomaly coefficient at UV and IR (see eg. \cite{anoms} for expressions in general higher curvature theories). In this construction, the starting point of the entropy functional played an important role in reaching the condition $16\alpha+5 \beta=0$. In \cite{ctheorems, ctheorems2}, this condition emerged by demanding the existence of a ``simple" c-function. Our analysis gives another way of looking at the same condition. One may wonder if it is possible to come up with a scenario where we did not need to absorb $E_{kk}$ into the c-function like what we have done. For example, instead of the HEE functional, perhaps one could try to construct a functional that cancels the $E_{kk}$ on the {\it rhs} of the Raychaudhuri equation. This does not appear to be the case--starting with the Wald entropy functional instead of the HEE functional does not naturally lead to the $16\alpha+5 \beta=0$ condition. 
 
%Note:- \par
%So we have shown that,
%\be
%\frac{d\Theta}{dt}\leq \kappa \theta.
%\ee
%$\kappa$ is poistivite. Now at the intial and the final equillibrium configuration i.e at $t=0$ and $t=\infty$ %$\Theta=0.$ So suppose, at some intermediate step $\Theta$ is negative, so 
%$\frac{d\Theta}{dt}$ will decrease it and $t=\infty$ $\Theta$ will be negative infinity. So there lies %the contradiction, hence $\Theta$ cannot be negative, it is always positive. So (23) proves that $\bar %\theta$ is positive. This sufficient for proving entropy change is positive as the integrand ( from (7) and %(8))  is positive for all values of $t$ from $0$  to $\infty$. 
%\par

%Now the change in entropy will be after integrating by parts from (7),
%\be
% S= t\,\, \theta\vert_{0}^{\infty}-\int d^3x \sqrt{h}dt (\frac{d\Theta}{dt}+\Theta\theta).
%\ee
%First term vanishes as at two asymptotic limit $\theta$ vanishes. Then using raychaudhuri equation one %can further establsh entropy increasing theorem from this. 
%\par
%So it proves the analog of area increase theorem for higher derivitive gravity. But of course GSL is not %used here. But there is some claim that with the null enery condition plus this one can show GSL ( that is %there in Sudipta +Wall previous papers,  allthough I do nnot proof of GSL  ). In general GSL is hard to %prove , here they are using simply the area increasing theorem. 
\section{Discussion}
In this paper, we considered four derivative gravity and subjected it to the condition that the second law of black hole thermodynamics is satisfied at linear and quadratic orders in perturbation. We showed that the entropy functional is uniquely fixed (upto second order in extrinsic curvatures) and coincides with the holographic entropy functionals. We also derived interesting bounds on the coupling which arises at next to leading order in perturbations.
We conclude here with a brief discussion on possible future directions. 
\begin{itemize}
\item One should consider more generic perturbations to see what happens at linear and second order. This is important since the possibility exists that for generic perturbations the second law holds only for very specific cases like the Lovelock theories. In \cite{camanho}, it was argued that causality constraints would require adding an infinite set of higher spin massive modes in a GB theory for consistency. It will be interesting to see if the second law knows about this in the perturbative sense used in the current work.

\item By considering asymptotically flat black holes and the Gauss-Bonnet theory, we found that for the negative GB coupling we get a minimum horizon radius for the second law to hold. Taking this to be a pathology, this appears to disfavour this sign of the coupling.
We further saw that the second law appears to know about the sound channel instablity in the context of Gauss-Bonnet holography. It will be interesting to see if $\lambda>-1/8$ which corresponds to the scalar channel instability \cite{Brigante} can also be seen from the second law.
\item One should consider how our analysis will change if there is matter coupling to the higher curvature terms. In this case, it is not clear how to define the null energy condition. It seems natural that the matter couplings will get constrained by demanding the second law.
\item Our analysis can be extended to general theories such as the quasi-topological theories \cite{quasitop}. In order to get a sufficient number of conditions to fix all the extrinsic curvature terms, one will need to turn on generic perturbations. 
\item In case of gravitational Chern-Simons terms, it will be interesting to see if and how the validity of second law can fix the entropy functionals \cite{loga}. Since the ${\it rhs}$ of the Raychaudhuri equation does not know about topological terms, the linearized analysis will not be able to capture the effect of such terms. One may need to consider topology changing processes like black hole mergers \cite{Sarkar:2010xp} to see the effect of such terms.
\item It will be interesting to compare our analysis with what arises from similar considerations in the fluid-gravity correspondence \cite{shiraz}. Demanding a positive entropy current for higher curvature theories will lead to constraints which are presumably going to be similar to what we have found.
\item The connection between the second law constraints and the positivity of relative entropy \cite{rel} in the context of holography can also be explored. We found that when we consider the second law for topological black holes with zero mass, we recover the bound from the tensor channel causality constraint. Why is this happening? A plausible explanation is as follows. The topological black holes with zero mass are just AdS spaces written in different co-ordinates and the finite Wald entropy of these black holes can be interpreted as an entanglement entropy across a sphere \cite{ctheorems, chm} in the dual field theory. Thus the second law in this context may be related to the positivity of relative entropy due to a time dependent perturbation corresponding to the infalling matter we have considered. Relative entropy is expected to be positive for a unitary field theory. Hence there appears to be an interesting interplay between bulk causality and field theory unitarity. It will be interesting to make this connection more precise.  It will also be interesting to understand the bounds in the context of c-theorems using the second law for causal horizons as advocated recently in \cite{sb}.

\end{itemize}

\section*{Acknowledgments} We thank Shamik Banerjee, Sayantani Bhattacharyya, Menika Sharma and Aron Wall for discussions. AB thanks IIT-Gandhinagar for hospitality during the course of this work. The research of SS is partially supported by IIT Gandhinagar start up grant No: IP/IITGN/PHY/SS/\\201415-12. AS acknowledges support from a Swarnajayanti fellowship, Govt. of India.

\end{document}